\documentstyle[prb,aps,psfig]{revtex}
\begin{document}

\title{
Pseudogap Phenomena in the BCS Pairing Model
}
\author{Satoshi {\sc Fujimoto}}
\address{
Department of Physics,
Kyoto University, Kyoto 606-8502, Japan
}
\date{\today}


\maketitle
\begin{abstract}
We investigate pseudogap phenomena realized in the
BCS pairing model with a long but finite interaction range.
We calculate the single-particle self-energy in all orders exactly
in the temperature range where the superconducting fluctuation
propagator is Gaussian-like. 
It is found that vertex corrections to the self-energy,
which are discarded in the previous studies, are crucially
important for the pseudogap of the single-particle density
of states in higher order calculations.
\end{abstract}

\pacs{PACS numbers: }




It is known that 
a pseudogap of the single-particle density of states (DOS) is
a universal phenomenon observed in electron systems interacting with
bosonic critical fluctuations.\cite{abra,maki,rice,sad,kampf,cas,seren,mck}
The interaction with a Gaussian fluctuation 
which arises near the instability toward 
a long-range order like antiferromagnetism, charge density wave, or 
superconductivity results in a much enhanced damping of single-particle
excitations,
leading to a decrease of the DOS at the Fermi level.
Recently, it has been proposed by some authors that
the pseudogap phenomena observed in high $T_c$ cuprates may be attributed to
the superconducting fluctuation.
\cite{rand,eme,gesh,tch,jan,engel,var,tremb,kyung,yanase,metz,tremb2} 
They have succeeded in explaining the pseudogap observed in ARPES, NMR, and
tunnel spectroscopy experiments.
In the previous studies,\cite{abra,maki,cas,tch,engel,var,tremb,yanase,metz} 
the single particle self-energy is calculated
by one-loop approximation ($t$-matrix approximation, diagrammatically
expressed as Fig.1(a)) or self-consistent
$t$-matrix approximation (expressed as Fig.1(b)).
Although these calculations show qualitative agreement with experimental
observations, there are some important differences between the results
obtained by different approximation methods.
For example, according to refs. 16 and 18, 
in the self-consistent one-loop approximation, 
the pseudogap of the DOS is destroyed near the Fermi level, and
the quasiparticle peak is restored.
On the other hand, such a restoration is not seen in the 
lowest-order one-loop approximation.\cite{yanase,metz}
Thus, it is desirable to examine the effects of vertex corrections,
which are expressed as Fig.1(c), and 
previously discarded without any justification of 
the approximation.
Moreover, the previous studies suggest that
the substantial decrease of the DOS at the Fermi level appears
for the interaction stronger than the intermediate coupling.
For this strong interaction, the validity of the $t$-matrix
approximation is obscure.
Motivated by these considerations, here,
we investigate a simple model for which we can calculate all-order
self-energy corrections 
in the temperature range where the superconducting
fluctuation is Gaussian-like.

\begin{figure}
\centerline{\psfig{file=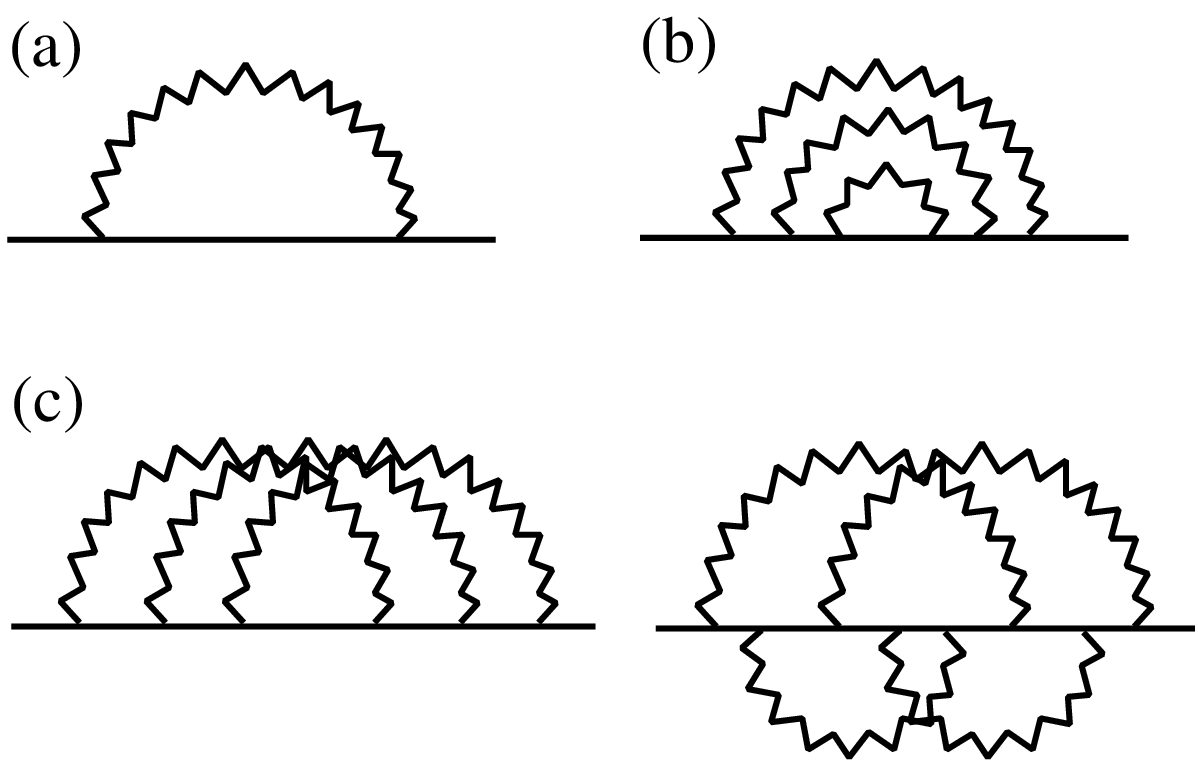,width=7cm}}
\caption{Diagrams for the irreducible single-particle self-energy.
The solid line represents the single electron Green's function.
The wavy line represents the propagator of the superconducting
fluctuation. (a) Lowest order one loop correction. (b) Self-consistent
one loop approximation. (c) Some typical diagrams of vertex corrections
which are not included in the diagrams (a) and (b).}
\end{figure}

The model Hamiltonian is given by
\begin{eqnarray}
&&H=\sum_{k,\sigma}E_kc^{\dagger}_{k\sigma}c_{k\sigma}
-\frac{1}{N}\sum_{q}V(q)B^{\dagger}(q)B(q), \label{ham} \\
&&V(q)=V\prod_{i=1}^d[\frac{1}{\pi}\frac{l_c}{1+l_c^2q^2_i}],
\end{eqnarray}
where $B(q)=\sum_k \xi_k c_{k\downarrow}c_{-k+q\uparrow}$, and
$d$ is the spatial dimension.\cite{yang}
The first term of eq.(\ref{ham}) is the kinetic energy 
with the energy dispersion $E_k$. 
The second term is the pairing attractive interaction with
the interaction range $l_c$.
$B(0)$ is a local annihilation operator of a Cooper pair with 
a structure factor $\xi_k$.
For example, $\xi_k=1$ for $s$-wave pairing, 
$\xi_k=\sin k_x,\sin k_y$ for $p$-wave
pairing, and $\xi_k=\cos k_x-\cos k_y, \cos k_x\cos k_y$ 
for $d$-wave pairing, and so forth.
In the limit of $l_c\rightarrow +\infty$, the second term of eq.(\ref{ham})
is reduced to $VB^{\dagger}(0)B(0)$.
Then eq.(\ref{ham}) is the pairing model,
which is exactly solvable.\cite{rich,cam,shas}
In this limit, the BCS mean field solution is exact, and 
the effects of fluctuation are completely suppressed.
For finite but sufficiently large $l_c$, the superconducting fluctuation
is restored, and can be calculated systematically using 
the expansion in terms of $1/l_c^d$.
We should bear in mind that although in this paper $l_c^d$ is taken 
as a large parameter,
$Vl_c^d\equiv U$ must be a finite constant even in the limit
of $l_c\rightarrow +\infty$ in order to keep the mean field transition
temperature finite, and $U$ is {\it not} a quantity of order $O(l_c^d)$. 

{\it Superconducting fluctuation propagator} --
We first discuss the propagator of the fluctuating bosonic field.
For this aim, following Krishnamurthy and Shastry,\cite{shas}
we introduce the auxiliary
boson field $\phi(q,\tau)$, and rewrite the interaction term of the action
as
\begin{eqnarray}
\sum_q\int^{\beta}_0d\tau\frac{V(q)}{N}B^{\dagger}(q,\tau)B(q,\tau)
&\rightarrow& \sum_q\int^{\beta}_0 d\tau \phi^{*}(q,\tau)\phi(q,\tau) \nonumber
\\
&-&{\rm i}\sum_q\sqrt{\frac{V(q)}{N}}\int^{\beta}_0d\tau[\phi^{*}(q,\tau)
B(q,\tau)+h.c.].
\end{eqnarray}
Integrating the fermion fields, we obtain the action for
the superconducting order parameter fluctuation.
\begin{eqnarray}
S&=&\sum_{q,m} (t+bq^2+a |\omega_m|)\phi^{*}_{q,m}\phi_{q,m} \nonumber  \\
&+&c\sum_{\stackrel{\scriptstyle m_1,m_2,m_3,}{k,k',q}}
\phi^{*}_{k'+q,m_2+m_3}\phi^{*}_{k-q,m_1-m_3}
\phi_{k,m_1}\phi_{k',m_2}+O((1/l_c^{d})^2). \label{bosact}
\end{eqnarray}
Here, $\phi_{q,m}$ is the Fourier transform of $\phi(q,\tau)$.
$t=(T-T_c)/T_c$, $a=3\zeta(2)N(0)U/(4\pi T_c)$,
 $b=7\zeta(3)N(0)Uv_F^2/(32\pi^2T_c^2)$, 
and $c=7\zeta(3)N(0)U/(16\pi^2l_c^dT_c^2)$ with $T_c$ the mean field 
transition temperature, and $N(0)$ the unrenormalized
DOS at the Fermi level. 
In the derivation of the quadratic term, we assumed that
the renormalization of the chemical potential caused by
the superconducting fluctuation is negligibly small.
In a two dimensional (2D) case, at sufficiently low temperatures,
the chemical potential may be strongly renormalized to a negative value, and
eventually, the crossover to the Bose-Einstein condensation
occurs at the zero temperature.\cite{bose}
In this paper, we do not consider such a low-temperature region
in the 2D case.
We also ignore the momentum and frequency dependencies of the coefficient
of the quartic term $c$. The momentum dependence is negligible,
since it is of a higher order in $1/l_c^d$. 
The energy dependence is also negligible in 
the low energy region $\omega \ll E_F$.
Note that $c$ is of the order $\sim 1/l_c^d$.
Thus, in the limit of $l_c \rightarrow +\infty$, 
the Gaussian term (the first term of eq.(\ref{bosact})) 
dominates, and the propagator of the boson field
is given by
\begin{equation}
\langle\phi_{q,n}\phi^{*}_{q,n}\rangle
=\frac{1}{a|\omega_n|+bq^2+t}. \label{gauss}
\end{equation}
This result was previously obtained by Krishnamurthy and Shastry.\cite{shas}
In this limit, the BCS mean field solution is exact, and
as a result, the single-particle self-energy correction vanishes.

To compute the effects of superconducting fluctuation on the self-energy,
we consider finite but very large $l_c$.
Then,  we need to take into account the effects of
the quartic and higher order terms of eq.(\ref{bosact}).
In the following, we consider only the quartic term, since it
gives leading corrections.
Applying the standard renormalization group method,\cite{reno} 
we obtain the one-loop scaling equations,
\begin{eqnarray}
\frac{d t}{dl}&=&2t+\frac{a_0 c}{1+t}, \label{sca1} \\
\frac{d c}{dl}&=&(2-d)c-\frac{b_0c^2}{(1+t)^2}. \label{sca2}
\end{eqnarray}
Here, $a_0$ and $b_0$ are positive constants.
From eq.(\ref{sca2}), we see that for $d \geq 2$, 
the quartic term is irrelevant or marginally irrelevant. 
The main effect of the quartic term is the renormalization of
the transition temperature, which is estimated from eq.(\ref{sca1}). 
The coefficients $a$ and $b$ of eq.(\ref{gauss}) are renormalized
by higher order corrections than two loop.
Thus, if the Ginzburg criterion for the smallness of fluctuations
is satisfied by the initial values of the parameters, $t$, $a$, $b$, and $c$,
the Gaussian-type propagator eq.(\ref{gauss}) with
the renormalized $t$, $a$, and $b$ can be regarded as the exact propagator 
of the superconducting fluctuation.
Now we consider the case of $ t\gg bl_c^{-2}$ in which, as will be shown
later, the all-order self-energy corrections can be calculated exactly.
In this case, the Ginzburg criterion is given by
\begin{equation}
\left(\frac{4cT_c}{l_c^d}\right)^{\frac{1}{2}} \ll t. \label{gc}
\end{equation}
Since $c\sim O(1/l_c^d)$, the condition eq.(\ref{gc}) is satisfied
when $l_c$ is sufficiently large to satisfy $t\gg bl_c^{-2}$.
In the following, we consider only the situation wherein
the unrenormalized coefficients $a$, $b$, $c$, and $t$ 
satisfy the Ginzburg criterion eq.(\ref{gc}),
and that eq.(\ref{gauss}) with the renormalized
parameters is a good approximation for the correlation function
of the superconducting fluctuation.

\begin{figure}
\centerline{\psfig{file=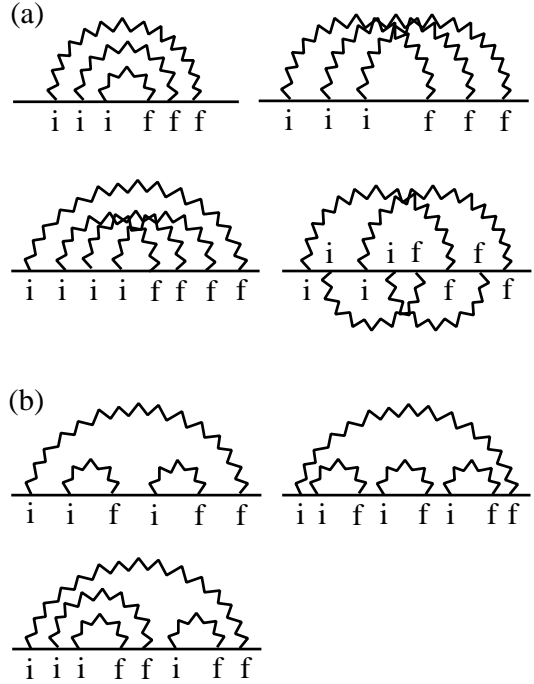,width=7cm}}
\caption{(a) Some typical self-energy diagrams 
in which all initial vertices are 
on the left of all final vertices. The symbol $i$ represents
the initial vertex, and $f$ the final vertex.
 (b) Some examples of diagrams which are not
included in $\Sigma^{(n)'}_k(\varepsilon)$. }
\end{figure}

{\it All-order self-energy corrections} --
We expand the irreducible single-particle self-energy in terms of $1/l_c^d$,
\begin{eqnarray}
\Sigma_k(\varepsilon)=\sum_{n=1}^{\infty} \Sigma_k^{(n)}(\varepsilon).
\end{eqnarray}
Here, $\Sigma^{(n)}_k(\varepsilon)$ is the $n$-th order term in $1/l_c^d$.
$\Sigma^{(n)}_k(\varepsilon)$ consists of two parts:
one part is the contribution from the diagrams
in which all initial vertices are on the left of all final vertices as shown
in Fig.2(a), and the other part is the contribution from
other irreducible diagrams as shown in Fig.2(b).
We express the former one as $\Sigma^{(n)'}_k(\varepsilon)$. 
Carrying out the frequency summation analytically, we obtain
\begin{eqnarray}
\Sigma^{(n)'}_k(\varepsilon)&=&\sum_{\{P\}}\frac{1}{N^n}
\sum_{q_1,...,q_n}\prod_{m=1}^{n}\frac{g_k(q_m)}{{\rm i}
\varepsilon+(-1)^{m-1}E_{k-\sum_{j=1}^m q_j}+{\rm i}\sum_{j=1}^m
C(q_j){\rm sgn}\varepsilon} \nonumber \\
&&\times\prod_{m'=1}^{n}\frac{1}{{\rm i}
\varepsilon+(-1)^{m'-1}E_{k-\sum_{j=1}^{m'} q_{P_j}}+{\rm i}\sum_{j=1}^{m'}
C(q_{P_j}){\rm sgn}\varepsilon}, \label{sig2}
\end{eqnarray}
where $g_k(q)=T\xi_k^2V(q)/(t+bq^2)$, and $C(q)=(t+bq^2)/a$.
$\{P\}\equiv\{P_1,P_2,...,P_n\}$ is generated by a permutation of 
$\{1,2,...,n\}$ under the rule that permutations between numbers with 
different parities are not allowed.  
For example, $\{P\}=\{1,2,3\},\{3,2,1\}$ for $n=3$, and
$\{P\}=\{1,2,3,4\},\{3,2,1,4\},\{1,4,3,2\},\{3,4,1,2\}$ for $n=4$, 
and so forth.
$\sum_{\{P\}}$ is the summation over these permutations.

In general, we cannot carry out $q$-summation of eq.(\ref{sig2})
analytically.
However, for $t \gg b/l_c^2$, eq.(\ref{sig2}) can be greatly simplified.
As we will see later, this case is very important for us.
In this limit, we can replace $q$-summation of eq.(\ref{sig2}),
$\frac{1}{N}\sum_q$, with $(1/l_c^d)\sum_q\delta_{q,0}$,
since the momentum transferred is restricted within $q<1/l_c\ll k_F$, 
for which $g_k(q)\approx g_k(0)$, $C(q)\approx C(0)$, and $E_{k-q}\approx E_k$.
In this approximation, all crossing diagrams of 
$\Sigma^{(n)'}_k(\varepsilon)$ give the same contribution
as its non-crossing diagrams.
This means that vertex corrections which are not included
in self-consistent $t$-matrix approximation are not negligible.
The total number of these diagrams, $N(n)$, is $m!(m+1)!$ for $n=2m+1$,
and $(m!)^2$ for $n=2m$.
Thus, we can rewrite eq.(\ref{sig2}) as,
\begin{eqnarray}
\Sigma^{(n)'}_k(\varepsilon)=N(n)\tilde{g}_k^n
\frac{1}{{\rm i}\varepsilon+(-1)^{n-1}E_k+{\rm i}nC(0){\rm sgn}\varepsilon}
\prod_{m=1}^{n-1}\frac{1}{({\rm i}\varepsilon+(-1)^{m-1}E_k+{\rm i}mC(0)
{\rm sgn}\varepsilon)^2},
\end{eqnarray}
where $\tilde{g}_k=g_k(0)/l_c^d$.
This expression has the same form as that obtained by Sadovskii
for one-dimensional fermion systems 
interacting with static Gaussian fluctuation.\cite{sad}
Then, we can apply Elyutin-Sadovskii's combinatorics method to evaluate
all diagrams.
According to Sadovskii, the contributions from 
$\Sigma^{(n)}_k(\varepsilon)-\Sigma^{(n)'}_k(\varepsilon)$ are expressed
in terms of $\Sigma^{(m)'}_k(\varepsilon)$ with $m<n$.\cite{sad}
Using Elyutin-Sadovskii's method, we can write 
the self-energy in all orders as a continued fraction form,
\begin{eqnarray}
\Sigma_k(\varepsilon)=\frac{\strut v(1)\tilde{g}_k}
{\displaystyle S_1(k,\varepsilon)
-\frac{\strut v(2)\tilde{g}_k}
{\displaystyle S_2(k,\varepsilon)
-\frac{\strut v(3)\tilde{g}_k}
{\displaystyle S_3(k,\varepsilon)
-\frac{\strut v(4)\tilde{g}_k}
{\displaystyle S_4(k,\varepsilon)-{}_{\ddots}}} }}, \label{self}
\end{eqnarray}
where $S_m(k,\varepsilon)=
{\rm i}\varepsilon+(-1)^{m-1}E_k+{\rm i}mC(0){\rm sgn}\varepsilon$, 
and $v(m)=[(m+1)/2]$, $[...]$ is Gauss's symbol. From this analytical
expression, we obtain the single-particle DOS,
$\rho(\varepsilon)=-1/\pi\sum_k {\rm Im}[1/(\varepsilon-E_k
-\Sigma_k^R(\varepsilon))]$.
We calculate $\rho(\varepsilon)$ numerically in the case of $s$-wave
pairing $\xi_k=1$, replacing $\sum_k$ with $N(0)\int d\varepsilon_k$. 
The result is shown in Fig.3(a).
The parameters are chosen so as to satisfy
the condition eq.(\ref{gc}) and $bl_c^{-2}\ll t$.
We see that the pseudogap appears near the Fermi level.
We also show the energy dependence of the self-energy in Figs.3(b) and 3(c).
It is seen that the scatterig rate, 
the imaginary part of the self-energy, is
much enhanced near the Fermi level, and that
the real part has a positive slope for $\varepsilon\sim 0$.
These features are characteristic of 
the pseudogap state.\cite{seren,mck,yanase} 
A similar result is obtained for other pairing symmetries, such as $p$-wave,
$d$-wave, and so forth. 

\begin{figure}
\centerline{\psfig{file=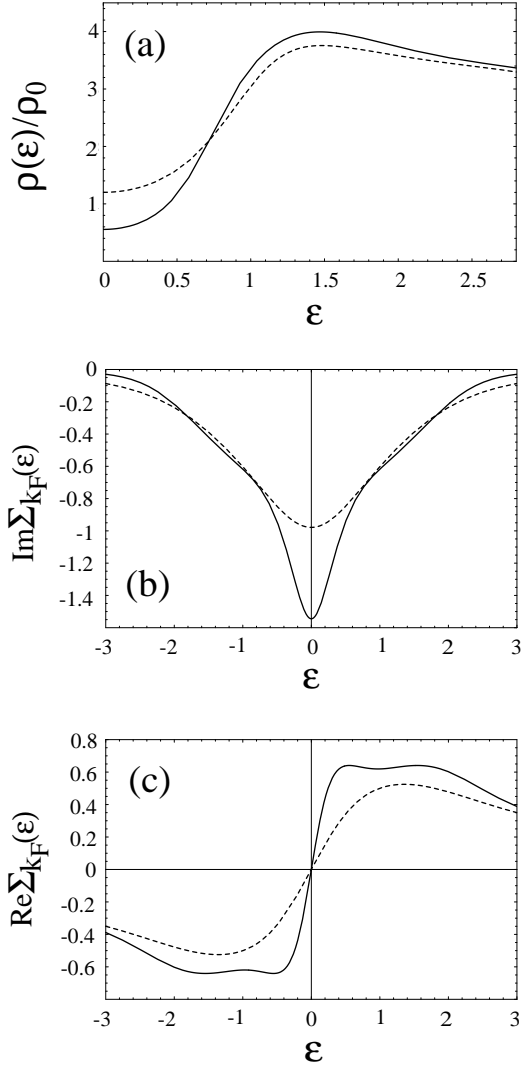,width=7cm}}
\caption{(a) The single particle DOS $\rho(\varepsilon)$ 
plotted as a function of energy. The vertical axis is renormalized
by the bare DOS $\rho_0$.
$\tilde{g}=1$, $C(0)=0.1$ for the solid line.
$\tilde{g}=1$, $C(0)=0.5$ for the dotted line.
(b) The imaginary part of the self-energy at the Fermi momentum
plotted as a function of energy.
The same parameters as Fig.3(a) are used.
(c) The real part of the self-energy at the Fermi momentum
plotted as a function of energy. The same parameters as above are used.}
\end{figure}
	
{\it Analytical results for single-particle damping}
-- The pseudogap behavior is due to 
the large enhancement of the damping,
which is caused by strong scattering with superconducting fluctuation.
Here, we will see how this enhancement is affected by the vertex corrections
to the self-energy which were discarded in the previous studies.
For this purpose, we calculate analytically
the damping of electrons at the Fermi level,
\begin{eqnarray}
\gamma\equiv -{\rm Im}\Sigma^R_{k_F}(0)=
\frac{\strut v(1)\tilde{g}_k}
{\displaystyle C(0)+\frac{\strut v(2)\tilde{g}_k}
{\displaystyle 2C(0)+\frac{\strut v(3)\tilde{g}_k}
{\displaystyle 3C(0)+\frac{\strut v(4)\tilde{g}_k}
{\displaystyle 4C(0)+{}_{\ddots}}} }}
\end{eqnarray}
The limiting value of this continued fraction is obtained from
the recursion relation, $P_{n+1}=(n+1)C(0)P_{n}+v(n+2)\tilde{g}P_{n-1}$,
$Q_{n+1}=(n+1)C(0)Q_n+v(n+2)\tilde{g}Q_{n-1}$, $P_0=0$, $Q_0=1$,
$P_{-1}=1$, $Q_{-1}=0$, and $\gamma
=\lim_{n\rightarrow\infty}P_n/Q_n$.\cite{frac} 
Up to a numerical constant, we have 
$\gamma\sim \tilde{g}_k/C(0)=aTU/(\pi^dl_c^dt^2)$.
For $\sqrt{aU/(\pi^dl_c^d)}\gg t$, the damping of electrons
is much enhanced, $\sim T/t^2\gg T$.
Since we consider the case of $t \gg b/l_c^2$, for which  the Ginzburg
criterion eq.(\ref{gc}) is also satisfied, 
the temperature range where the above result is applicable is
$b/l_c^2 \ll t \ll \sqrt{aU/(\pi^dl_c^d)}$.
It is noted that the result of all-order calculation gives
the same asymptotic behavior as that obtained by
the lowest-order one-loop approximation which is also given by
$\sim \tilde{g}/C(0)\sim T/t^2$.
This implies that higher order corrections cancel each
other, and the lowest-order term gives the leading contribution.
This cancellation is not due to the presence of the Ward identity.
We do not have the Ward identity which relates a self-energy to a vertex
in particle-particle channel.
Actually, the cancellation takes place among different order diagrams.
The total number of the diagrams which give dominant contribution in the $n$-th
order increases as $O((n!)^2)$. As a result,
higher order contributions cancel with lower order contributions
effectively.
We compare this result with that obtained from self-consistent
$t$-matrix approximation.
The self-energy in the self-consistent approximation is derived by
substituting $v(m)=1$ for all $m$ of eq.(\ref{self}).
The damping at the Fermi level, $\gamma_{sc}$, is given by
\begin{eqnarray}
\gamma_{sc}=
\frac{\strut \tilde{g}_k}
{\displaystyle C(0)+\frac{\strut \tilde{g}_k}
{\displaystyle 2C(0)+\frac{\strut \tilde{g}_k}
{\displaystyle 3C(0)+\frac{\strut \tilde{g}_k}
{\displaystyle 4C(0)+{}_{\ddots}}} }}=\frac{{\rm i}\sqrt{\tilde{g}_k}
J_1({\rm i}\frac{2\sqrt{\tilde{g}_k}}{C(0)})}{J_0({\rm i}
\frac{2\sqrt{\tilde{g}_k}}{C(0)})}, \label{damp2}
\end{eqnarray}
where $J_n(x)$ is the $n$-th Bessel function.
For sufficiently small $t$ (large $\sqrt{\tilde{g}_k}/C(0)$),
eq.(\ref{damp2}) is well approximated by 
$\sqrt{\tilde{g}_k}=[\xi_k^2UT/(tl_c^d\pi^d)]^{1/2}\sim 1/\sqrt{t}$.
This is much suppressed compared with the result of all-order calculations.
For $t\gg b/l_c^2$, $\sqrt{\tilde{g}_k} \ll 
[\xi_k^2UT/(\pi^dbl_c^{d-2})]^{1/2}$.
Thus, in the case of $d=2$ or $3$, the damping $\gamma_{sc}$ 
cannot be enhanced sufficiently under this condition.
This result presents a clear contrast with the all-order result.
The self-consistent $t$-matrix approximation, 
which neglects vertex corrections underestimates crucially the effects of
strong scattering by critical fluctuations, and may 
cause a fatal error not only quantitatively but also qualitatively 
in a pseudogap state. 
These results instructively show that the lowest-order one-loop
calculation is more reliable than
the self-consistent one-loop treatment, and furthermore imply that
the lowest order one-loop treatment of the self-energy corrections
may be a fairly good approximation
because of the cancellation between higher order corrections.

In summary, we have studied the pseudogap of the
single particle DOS in the BCS pairing model
with a long but finite interaction range using a well controlled
approximation method.
We have derived the all-order single-particle self-energy exactly
in the temperature range where the superconducting fluctuation propagator
is a Gaussian form.
It has been revealed that the vertex corrections to the self-energy
are essentially
important in higher order calculations.

The author would like to thank K. Yamada and 
B. S. Shastry for useful conversations.
This work was partly supported by a Grant-in-Aid from the Ministry
of Education, Culture, Science, and Technology, Japan. 



                                                                    

\begin{references}
\bibitem{abra} E. Abraham, M Redi, and C. Woo: Phys. Rev. B{\bf 1}
(1970) 218.

\bibitem{maki} J. P. Hurault and K. Maki: Phys. Rev. B{\bf 2} (1970)
2560.

\bibitem{rice} M. J. Rice and S. Str\"assler: Solid State Communication
{\bf 13} (1973) 1389.

\bibitem{sad} M. V. Sadovskii: Sov. Phys. JETP {\bf 50} (1979) 989.

\bibitem{kampf} A. Kampf and J. R. Schrieffer: Phys. Rev. B{\bf 41}
(1990) 6399.

\bibitem{cas} C. Di Castro, et al.:
Phys. Rev. B{\bf 49} (1990) 10211.

\bibitem{seren} J. J. Deisz, D. W. Hess, and J. W. Serene: Phys. Rev. Lett.
{\bf 76} (1996) 1312.

\bibitem{mck} R. H. McKenzie and D. Scarratt: Phys. Rev. B{\bf 54} (1996)
12709.

\bibitem{rand} M. Randeria et al.: Phys. Rev. Lett. {\bf 69} (1992)
2001; M. Randeria and A. A. Varlamov: Phys. Rev. B{\bf 50} (1994)
10401.

\bibitem{eme} V. Emery and S. Kivelson: Nature {\bf 374} (1995)
434.

\bibitem{gesh} V. B. Geshkenbein, L. B. Ioffe, and A. I. Larkin:
Phys. Rev. B{\bf 55} (1997) 3173.

\bibitem{tch} O. Tchernyshyov: Phys. Rev. B{\bf 56} (1997)
3372.

\bibitem{jan} B. Janko, J. Maly, and K. Levin: Phys. Rev. B{\bf 56}
(1997) 11407.

\bibitem{engel} J. R. Engelbrecht et al.: Phys. Rev. B{\bf 57}
(1998) 13406.

\bibitem{var} A. A. Varlamov et al.: Adv. Phys. {\bf 48}
(1999) 655.

\bibitem{tremb} S. Moukouri et al.: Phys. Rev. B{\bf 61} (2000) 7887;

\bibitem{kyung} B. Kyung: Phys. Rev. B{\bf 63} (2000) 014502.

\bibitem{yanase} Y. Yanase and K. Yamada: J. Phys. Soc. Jpn. {\bf 70}
(2001) 1659.

\bibitem{metz} D. Rohe and W. Metzner: Phys. Rev. B{\bf 63} (2001) 224509.

\bibitem{tremb2} B. Kyung, S. Allen, and A.-M. S. Tremblay: 
Phys. Rev. B{\bf 64} (2001) 075116. The importance of vertex corrections
for the pseudogap phenomena is also pointed out in this paper.

\bibitem{yang} A simlar but different model was also studied 
in K. Yang and S. L. Sondhi, Phys. Rev. B{\bf 62} (2000) 11778.

\bibitem{rich} R. W. Richardson: Phys. Lett. {\bf 3} (1963) 277;
J. Math. Phys. {\bf 18} (1977) 1802.

\bibitem{cam} M. C. Camiaggio, A. M. F. Rivas, and M. Saraceno:
Nucl. Phys. A{\bf 624} (1997) 157.

\bibitem{shas} H. R. Krishnamurthy and B. S. Shastry: Phys. Rev. Lett.
{\bf 84} (2000) 4918.

\bibitem{bose} S. Schmitt-Rink, C. M. Varma, and A. E. Ruckenstein:
Phys. Rev. Lett. {\bf 63} (1989) 445.

\bibitem{reno} For example, J. Zinn-Justin: {\it Quantum Field Theory
and Critical Phenomena} (Oxford, New York, 1993);
R. Shankar: Rev. Mod. Phys. {\bf 66} (1994) 129.

\bibitem{frac} H. S. Wall:
{\it Analytic Theory of Continued Fractions}
(Chelsea, New York, 1967).


\end{references}
\end{document}